\documentclass[11pt]{article}
\usepackage{a4p}
\usepackage{amssymb}
\usepackage{epsfig}
\usepackage{cite}
\usepackage{pennames}
\usepackage{rotating}
\newcommand{\intLdt}  {57.21}

\newcommand{\dLstat}  {0.15}
\newcommand{\dLsys}   {0.20}

\newcommand{\rroots}  {182.68}  
\def\mrm{\mathrm}
\newcommand{\epem}{\ensuremath{\mathrm{e}^+\mathrm{e}^-}}

\newcommand{\lplm}{\ensuremath{\ell^+\ell^-}}
\newcommand{\Zz}{\ensuremath{{\mathrm{Z}^0}}}
\newcommand{\WW}{\ensuremath{\mathrm{W}^+\mathrm{W}^-}}

\newcommand{\qq}{\ensuremath{\mathrm{q\overline{q}}}}

\newcommand{\lnu}{\ensuremath{\ell\overline{\nu}_{\ell}}}
\newcommand{\lpnu}{\ensuremath{\ell^+ \nu_{\ell}}}
\newcommand{\lmnu}{\ensuremath{{\ell^{\prime}}^-\overline{\nu}_{\ell^{\prime}}}}
\newcommand{\enu}{\ensuremath{\mathrm{e\overline{\nu}_{e}}}}
\newcommand{\mnu}{\ensuremath{\mu\overline{\nu}_{\mu}}}
\newcommand{\tnu}{\ensuremath{\tau\overline{\nu}_{\tau}}}

\newcommand{\qqqq}{\ensuremath{\qq\qq}}
\newcommand{\qqln}{\ensuremath{\qq\lnu}}

\newcommand{\WWqqln}{\ensuremath{\WW\rightarrow\qq\lnu}}
\newcommand{\WWqqqq}{\ensuremath{\WW\rightarrow\qq\qq}}
\newcommand{\WWqqen}{\ensuremath{\WW\rightarrow\qq\enu}}
\newcommand{\WWqqmn}{\ensuremath{\WW\rightarrow\qq\mnu}}
\newcommand{\WWqqtn}{\ensuremath{\WW\rightarrow\qq\tnu}}

\newcommand{\WWlnln}{\ensuremath{\WW\rightarrow\lpnu\lmnu}}

\newcommand{\Wenu}{\ensuremath{\epem \rightarrow \mathrm{W}\enu}}

\newcommand{\ZZqqll}{\ensuremath{(\Zz/\gamma)^*(\Zz/\gamma)^*\rightarrow\qq\lplm}}

\newcommand{\Zqq}{\ensuremath{\Zz/\gamma\rightarrow\qq}}

\newcommand{\Mz}{\ensuremath{M_{\mrm{Z}}}}
\newcommand{\Mw}{\ensuremath{M_{\mrm{W}}}}
\newcommand{\Mwqqqq}{\ensuremath{M_{\mrm{W}}^{\qq\qq}}}
\newcommand{\Mwqqln}{\ensuremath{M_{\mrm{W}}^{\qq\lnu}}}
\newcommand{\Gw}{\ensuremath{\Gamma_{\mathrm{W}}}}
\newcommand{\DMw}{\ensuremath{\Delta M_{\mrm{W}}}}

\newcommand{\Opal}{\mbox{O{\sc pal}}}
\newcommand{\Delphi}{\mbox{D{\sc elphi}}}
\newcommand{\Lep}{\mbox{L{\sc ep}}}

\newcommand{\Jetset}{\mbox{J{\sc etset}}}
\newcommand{\Koralw}{\mbox{K{\sc oralw}}}
\newcommand{\KORALW}{\mbox{K{\sc oralw}}}

\newcommand{\Excalibur}{\mbox{E{\sc xcalibur}}}
\newcommand{\EXCALIBUR}{\mbox{E{\sc xcalibur}}}
\newcommand{\grcff}{\mbox{grc4f}}
\newcommand{\Pythia}{\mbox{P{\sc ythia}}}
\newcommand{\PYTHIA}{\mbox{P{\sc ythia}}}
\newcommand{\Ariadne}{\mbox{A{\sc riadne}}}

\newcommand{\Herwig}{\mbox{H{\sc erwig}}}
\newcommand{\HERWIG}{\mbox{H{\sc erwig}}}

\newcommand{\GeV}{\ensuremath{\mathrm{GeV}}}

\newcommand{\roots}{\ensuremath{\sqrt{s}}}

\newcommand{\Zgamma}{\ensuremath{\Zz/\gamma}}

\newcommand {\ra}         {\ensuremath{\rightarrow}}

\newcommand{\CC}{\mbox{{\sc CC03}}}
\newcommand{\mrec}{\ensuremath{m_{\mathrm{rec}}}}
\newcommand{\srec}{\ensuremath{\sigma_{\mathrm{rec}}}}
\def\etal{\mbox{{\it et al.}}}
\def\ie{\mbox{{\it i.e.}}}
\def\eg{\mbox{{\it e.g.}}}

\def\gappeq{\ensuremath{\mathrel{ \rlap{\raise.5ex\hbox{>}}
                      {\lower.5ex\hbox{\sim}}}}}
\def\lappeq{\ensuremath{\mathrel{ \rlap{\raise.5ex\hbox{<}}
                      {\lower.5ex\hbox{\sim}}}}}

\newcommand{\JPG}[3]  {J.\ Phys.\ \textbf{G#1} (#2) #3}
\newcommand{\PLB}[3]  {Phys.\ Lett.\ \textbf{B#1} (#2) #3}
\newcommand{\ZPC}[3]  {Z.\ Phys.\ \textbf{C#1} (#2) #3}
\newcommand{\EPC}[3]  {Eur.\ Phys.\ J.\ \textbf{C#1} (#2) #3}
\newcommand{\NIMA}[3] {Nucl.\ Instr.\ Meth.\ \textbf{A#1} (#2) #3}

\newcommand{\PRL}[3]  {Phys.\ Rev.\ Lett.\ \textbf{#1} (#2) #3}
\newcommand{\PRD}[3]  {Phys.\ Rev.\ \textbf{D#1} (#2) #3}
\newcommand{\NPB}[3]  {Nucl.\ Phys.\ \textbf{B#1} (#2) #3}

\newcommand{\CPC}[3]  {Comput.\ Phys.\ Commun.\ \textbf{#1} (#2) #3}
\newcommand{\WWqqenu}{\mbox{\WW$\rightarrow$ \qq\enu}}
\newcommand{\WWqqmnu}{\mbox{\WW$\rightarrow$ \qq\mnu}}
\newcommand{\WWqqtnu}{\mbox{\WW$\rightarrow$ \qq\tnu}}

\newcommand{\Leptwo}{\mbox{L{\sc EP2}}}

\def\opalabbiendi{OPAL Collaboration, G.\ Abbiendi \etal}
\def\opalackerstaff{OPAL Collaboration, K.\ Ackerstaff \etal}
\def\opalalexander{OPAL Collaboration, G.\ Alexander \etal}

\def\opalakrawy{OPAL Collaboration, M.Z.\ Akrawy \etal}

\begin{document}
\begin{titlepage}
\begin{center}
  {\Large   EUROPEAN LABORATORY FOR PARTICLE PHYSICS}
\end{center}\bigskip
\begin{flushright}
  CERN-EP/98-197 \\  
  December 10, 1998
\end{flushright}
\bigskip\bigskip\bigskip
\begin{center}
  {\huge\bf \boldmath Measurement of the W Mass and Width \\
    \vspace*{5mm}    
    in \epem\ Collisions at 183~GeV}
\end{center}
\bigskip\bigskip
\begin{center}
  {\LARGE The OPAL Collaboration}
\end{center}\bigskip\bigskip 
%
% --- pigpigpig
%
\begin{center}
  {\large \bf Abstract }
\end{center} 
Using a data sample of 57~pb$^{-1}$ recorded at a centre-of-mass energy of 
183~GeV with the \Opal\ detector at \Lep, 282 \WWqqqq\ and 
300 \WWqqln\ candidate events are used to obtain a measurement of the mass of 
the W boson, 
{\ensuremath{\Mw=80.39\pm0.13\mrm{(stat.)}\pm0.05\mrm{(syst.)}}}~GeV,
assuming the Standard Model relation between \Mw\ and \Gw.  A second fit
provides a direct measure of the width of the W boson and gives
{\ensuremath{\Gw=1.96\pm0.34 \mrm{(stat.)}\pm0.20\mrm{(syst.)}}}~GeV. These
results are combined with previous \Opal\ results 
to obtain {\ensuremath{\Mw=80.38\pm0.12\mrm{(stat.)}\pm0.05\mrm{(syst.)}}}~GeV
and {\ensuremath{\Gw=1.84\pm0.32\mrm{(stat.)}\pm0.20\mrm{(syst.)}}}~GeV.
\vspace*{1cm}

\begin{center}
  {\large 
    (submitted to Physics Letters B) }
\end{center}

\end{titlepage}

% --- Author List
\begin{center}{\Large        The OPAL Collaboration
}\end{center}\bigskip
\begin{center}{
%begin authorlist PLEASE DO NOT DELETE THIS COMMENT
G.\thinspace Abbiendi$^{  2}$,
K.\thinspace Ackerstaff$^{  8}$,
G.\thinspace Alexander$^{ 23}$,
J.\thinspace Allison$^{ 16}$,
N.\thinspace Altekamp$^{  5}$,
K.J.\thinspace Anderson$^{  9}$,
S.\thinspace Anderson$^{ 12}$,
S.\thinspace Arcelli$^{ 17}$,
S.\thinspace Asai$^{ 24}$,
S.F.\thinspace Ashby$^{  1}$
% ...
%end authorlist PLEASE DO NOT DELETE THIS COMMENT
}\end{center}\bigskip
\bigskip
%begin institutes
$^{  1}$School of Physics and Astronomy, University of Birmingham,
Birmingham B15 2TT, UK
\newline
$^{  2}$Dipartimento di Fisica dell' Universit\`a di Bologna and INFN,
I-40126 Bologna, Italy
\newline
$^{  3}$Physikalisches Institut, Universit\"at Bonn,
D-53115 Bonn, Germany
\newline
$^{  4}$Department of Physics, University of California,
Riverside CA 92521, USA
\newline
$^{  5}$Cavendish Laboratory, Cambridge CB3 0HE, UK
\newline
$^{  6}$Ottawa-Carleton Institute for Physics,
Department of Physics, Carleton University,
Ottawa, Ontario K1S 5B6, Canada
\newline
$^{  7}$Centre for Research in Particle Physics,
Carleton University, Ottawa, Ontario K1S 5B6, Canada
\newline
$^{  8}$CERN, European Organisation for Particle Physics,
CH-1211 Geneva 23, Switzerland
\newline
$^{  9}$Enrico Fermi Institute and Department of Physics,
University of Chicago, Chicago IL 60637, USA
\newline
$^{ 10}$Fakult\"at f\"ur Physik, Albert Ludwigs Universit\"at,
D-79104 Freiburg, Germany
\newline
$^{ 11}$Physikalisches Institut, Universit\"at
Heidelberg, D-69120 Heidelberg, Germany
\newline
$^{ 12}$Indiana University, Department of Physics,
Swain Hall West 117, Bloomington IN 47405, USA
\newline
$^{ 13}$Queen Mary and Westfield College, University of London,
London E1 4NS, UK
\newline
$^{ 14}$Technische Hochschule Aachen, III Physikalisches Institut,
Sommerfeldstrasse 26-28, D-52056 Aachen, Germany
\newline
$^{ 15}$University College London, London WC1E 6BT, UK
\newline
$^{ 16}$Department of Physics, Schuster Laboratory, The University,
Manchester M13 9PL, UK
\newline
$^{ 17}$Department of Physics, University of Maryland,
College Park, MD 20742, USA
\newline
$^{ 18}$Laboratoire de Physique Nucl\'eaire, Universit\'e de Montr\'eal,
Montr\'eal, Quebec H3C 3J7, Canada
\newline
$^{ 19}$University of Oregon, Department of Physics, Eugene
OR 97403, USA
\newline
$^{ 20}$CLRC Rutherford Appleton Laboratory, Chilton,
Didcot, Oxfordshire OX11 0QX, UK
\newline
$^{ 22}$Department of Physics, Technion-Israel Institute of
Technology, Haifa 32000, Israel
\newline
$^{ 23}$Department of Physics and Astronomy, Tel Aviv University,
Tel Aviv 69978, Israel
\newline
$^{ 24}$International Centre for Elementary Particle Physics and
Department of Physics, University of Tokyo, Tokyo 113-0033, and
Kobe University, Kobe 657-8501, Japan
\newline
$^{ 25}$Institute of Physical and Environmental Sciences,
Brunel University, Uxbridge, Middlesex UB8 3PH, UK
\newline
$^{ 26}$Particle Physics Department, Weizmann Institute of Science,
Rehovot 76100, Israel
\newline
$^{ 27}$Universit\"at Hamburg/DESY, II Institut f\"ur Experimental
Physik, Notkestrasse 85, D-22607 Hamburg, Germany
\newline
$^{ 28}$University of Victoria, Department of Physics, P O Box 3055,
Victoria BC V8W 3P6, Canada
\newline
$^{ 29}$University of British Columbia, Department of Physics,
Vancouver BC V6T 1Z1, Canada
\newline
$^{ 30}$University of Alberta,  Department of Physics,
Edmonton AB T6G 2J1, Canada
\newline
$^{ 31}$Duke University, Dept of Physics,
Durham, NC 27708-0305, USA
\newline
$^{ 32}$Research Institute for Particle and Nuclear Physics,
H-1525 Budapest, P O  Box 49, Hungary
\newline
$^{ 33}$Institute of Nuclear Research,
H-4001 Debrecen, P O  Box 51, Hungary
\newline
$^{ 34}$Ludwigs-Maximilians-Universit\"at M\"unchen,
Sektion Physik, Am Coulombwall 1, D-85748 Garching, Germany
\newline
%end institutes
\bigskip\newline
%begin notes
$^{  a}$ and at TRIUMF, Vancouver, Canada V6T 2A3
\newline
$^{  b}$ and Royal Society University Research Fellow
\newline
$^{  c}$ and Institute of Nuclear Research, Debrecen, Hungary
\newline
$^{  d}$ and Department of Experimental Physics, Lajos Kossuth
University, Debrecen, Hungary
\newline
$^{  e}$ on leave of absence from the University of Freiburg
\newline
%end notes

% --- Introduction
\newpage

\section{Introduction}
\label{intro} 

Comparison between direct measurements of the mass of the W boson, \Mw, and the
value determined indirectly from precise electroweak results from data taken at
$\roots\approx\Mz$ and lower energies \cite{bib:lep-ewksum} provides an 
important test of the Standard Model. In addition, the direct measurement of 
\Mw\ can be used to constrain the mass of the Higgs boson, $M_{\mrm{H}}$, by 
comparison with theoretical predictions involving radiative corrections 
sensitive to $M_{\mrm{H}}$ \cite{bib:sirlin}.  The constraints imposed using 
\Mw, which are presently limited by statistical uncertainties, are 
complementary to those imposed by other electroweak measurements, which are 
largely limited by theoretical uncertainties~\cite{bib:degressi}.

The first direct measurements of the W boson mass were performed at hadron 
colliders \cite{bib:HAD-mw}.  During 1996, the \Lep\ collider at CERN began 
operating at centre-of-mass energies exceeding the \WW\ production 
threshold (\Leptwo) thus allowing direct determinations of the mass of the W 
boson.  The combination of direct measurements from \Leptwo\ [5-9] and
from hadron colliders
%\cite{bib:O-mw172}--$\!\!$\cite{bib:ADL-mw161} and from hadron colliders 
presently yields $\Mw = 80.41\pm 0.10$~GeV \cite{bib:pdg98}.  It is expected 
that \Leptwo\ will ultimately achieve a precision on the W mass of 
approximately $30$-$40$~MeV \cite{bib:LEP2YR}.

In 1997, \Opal\ collected  $57$~pb$^{-1}$ of data at a centre-of-mass energy of
approximately $183$~GeV. This paper describes measurements of the W boson mass
and width using this data sample.  

% --- BEGIN... 

\section{The OPAL Detector} 
\label{opaldet}

The \Opal\ detector includes a 3.7 m diameter tracking volume immersed in a 
$0.435$ T axial magnetic field, which yields a transverse\footnote{The \Opal\ 
coordinate system is defined such that the $z$-axis is parallel to and in the 
direction of the $\mrm{e}^{-}$ beam, the $x$-axis lies in the plane of the 
accelerator pointing towards the centre of the \Lep\ ring, and the $y$-axis is
normal to the plane of the accelerator and has its positive direction defined 
to yield a right-handed coordinate system.  The azimuthal angle, $\phi$, and 
the polar angle, $\theta$, are the conventional spherical coordinates.} 
momentum resolution of $\sigma_{p_{xy}}/p_{xy} \approx
  \sqrt{(0.020)^{2} + (0.0015 \cdot p_{xy})^{2}/\mrm{GeV}^{2}}$ and an average
angular resolution of about $0.3$ mrad in $\phi$ and $1$ mrad in $\theta$. The
electromagnetic calorimeter consists of $11\,704$ lead glass blocks with full 
azimuthal acceptance in the range $\left| \cos\theta \right| < 0.98$ and a 
relative energy resolution of approximately $3\%$ at $E\approx47$~GeV, the mean
energy of electrons from W decays.  The magnet return yoke is instrumented with
streamer tubes to serve as the hadronic calorimeter.  Muon chambers surrounding
the hadronic calorimeter provide muon identification over the range 
$\left| \cos\theta \right| < 0.98$. Jets are constructed from charged tracks 
and energy deposits in the electromagnetic and hadronic calorimeters using the
Durham algorithm \cite{bib:durham}.  The energies of reconstructed jets are 
calculated using the technique described in~\cite{bib:GCE}.  Using 
$\epem\ra\qq$ data taken at $\roots=91$~GeV, the jet energy resolution is 
determined to be approximately $\sigma_{E}/E \approx 20\%$ with an angular 
resolution of $20$-$30$~mrad depending on the jet visible energy and polar 
angle.  A more detailed description of the \Opal\ detector can be found in 
\cite{bib:opaldet}.

% --- data and MC samples

\section{Data and Monte Carlo Samples}
\label{data-mc}

The integrated luminosity of the data sample, evaluated using small
angle Bhabha scattering events observed in the silicon tungsten forward 
calorimeter, is 
$\intLdt\pm\dLstat \mrm{(stat.)} \pm\dLsys \mrm{(syst.)}$~pb$^{-1}$.  The 
mean centre-of-mass energy, weighted by luminosity, is 
$\roots= \rroots \pm 0.05$~\GeV~\cite{bib:energy183}.

\subsection{Event Selections}
\label{sec:evtsel}

The event selections are described in detail in~\cite{bib:O-tgc183}.  The 
selections are sensitive to the leptonic \WWlnln, semi-leptonic \WWqqln\ and 
hadronic \WWqqqq\ final states.  By construction, the selections are mutually 
exclusive.  The leptonic final state is not used in this analysis.

Semi-leptonic \WWqqln\ decays comprise $44\%$ of the total \WW\ cross-section.
The selection employs three multivariate relative likelihood discriminants, one
for each of the \WWqqen, \WWqqmn\ and \WWqqtn\ final states.  The \WWqqen\ and
\WWqqmn\ channels are characterized by two well-separated hadronic jets, a 
high-momentum lepton and missing momentum due to the prompt neutrino from the 
leptonic W decay.  The \WWqqtn\ channel is characterized similarly except that
the $\tau$ lepton is identified as an isolated, low-multiplicity jet typically
consisting of one or three charged tracks.  After all cuts, \WWqqln\ events are
selected with an efficiency of $85\%$ and a purity of $90\%$.  The dominant 
backgrounds are \Zqq\ and four-fermion processes such as \Wenu\ and \ZZqqll.

Hadronic \WWqqqq\ decays comprise $46\%$ of the total \WW\ cross-section and 
are characterized by four energetic hadronic jets and little or no missing 
energy. A loose preselection removes approximately $98\%$ of the dominant
background process, \Zqq .  Following the preselection, a multivariate relative
likelihood discriminant is employed to select the \WWqqqq\ candidates with an 
efficiency of $85\%$ and a purity of $78\%$.

After these selections, 361 \WWqqln\ and 438 \WWqqqq\ candidate events are 
identified, consistent with Standard Model expectations \cite{bib:O-tgc183}.  
As discussed in Section~\ref{mrec}, additional cuts are applied to remove 
poorly reconstructed events and further reduce backgrounds.

\subsection{Monte Carlo Samples} 
\label{sec:mcsamples}

A number of Monte Carlo simulation programs are used to provide estimates of 
efficiencies and purities as well as the shapes of the W mass distributions.
The majority of the samples are generated at \roots\ = 183~GeV assuming
$\Mw=80.33$~\GeV. All Monte Carlo samples include a full simulation of the 
\Opal\ detector~\cite{bib:GOPAL}. 

The main \WW\ samples are generated using \KORALW \cite{bib:KORALW} and include
only the \CC\ diagrams\footnote{In this paper, the doubly-resonant W pair 
production diagrams, {\em i.e.} $t$-channel $\nu_{\mathrm{e}}$ exchange and 
$s$-channel \Zgamma\ exchange, are referred to as ``\CC'', following the 
notation of \cite{bib:LEP2YR}.}.  The four-fermion backgrounds 
$\mrm{We}\overline{\nu}_e$ and $(\Zz/\gamma)^*(\Zz/\gamma)^*$ are simulated 
using \Pythia\ \cite{bib:PYTHIA}, while \grcff\ \cite{bib:GRC4F} and 
\EXCALIBUR\ \cite{bib:EXCALIBUR} are used to estimate systematic uncertainties.
The background process \Zqq\ is simulated using \PYTHIA, with \HERWIG\ 
\cite{bib:HERWIG} used as an alternative to assess possible systematic effects.

% --- W mass

\section{Measurement of the Mass and Width of the W Boson}
\label{Wmass}

The W boson mass, \Mw, and decay width, \Gw, are determined from fits to the
reconstructed invariant mass spectrum of the selected \WWqqln\ and \WWqqqq\ 
events. For each selected event, a kinematic fit is employed to improve the 
mass resolution and further reduce background.  A reweighting technique 
\cite{bib:LEP2YR} is used to produce Monte Carlo mass spectra corresponding to
any given mass and width.  A binned likelihood fit is used to determine \Mw\ 
and \Gw\ by comparing the shape of the reconstructed invariant mass 
distribution from the data to that from reweighted Monte Carlo spectra.

Two alternative methods are also used to extract \Mw .  In the first, an 
analytic fit to the measured mass spectrum uses an unbinned likelihood fit to
determine \Mw .  To describe the signal shape, the fit uses a parametrization 
based on a Breit-Wigner function \cite{bib:O-mw172}.  The second method uses a
convolution technique similar to that used by the \Delphi\ Collaboration 
\cite{bib:D-mw172}.  These alternative fits have sensitivities similar to that
of the reweighting fit and are used as cross-checks.

\subsection{Invariant Mass Reconstruction}
\label{mrec}

The three methods for extracting \Mw\ use essentially the same procedures
to reconstruct the invariant mass of the W candidates.  The description
provided here applies to the reweighting method.  Small variations relevant
for the alternative analyses are discussed in Section~\ref{sec:alt-fits}.

For the selected \WWqqqq\ events tracks and clusters are grouped into four jets
using the Durham algorithm. A kinematic fit is then performed to estimate the 
reconstructed invariant mass of the W candidate.  The fit incorporates the 
constraints of energy and momentum conservation (4C fit) yielding two 
reconstructed masses per event, one for each W boson in the final state. An 
additional constraint can be incorporated by neglecting the finite W width and
constraining the masses of the two W boson candidates to be equal (5C fit), 
thus yielding a single reconstructed invariant mass for each event. The 
kinematic fit employs the method of Lagrange multipliers and a 
$\chi^2$-minimization technique.  For extracting \Mw\ from the \WWqqqq\ 
candidates, the 5C fit is used to determine a reconstructed invariant mass, 
\mrec, its error, \srec, and a $\chi^2$ fit-probability for each event.  The 
measured jet momenta with their associated errors and the measured jet masses 
are used as inputs.  The use of the measured jet masses, rather than treating 
the jets as massless, improves the fitted mass resolution. Based on Monte Carlo
studies, the errors associated with the measured jet momentum are parameterized
as functions of the visible energy and polar angle of the jet.

\begin{figure}
  \epsfxsize=\textwidth
  \epsffile{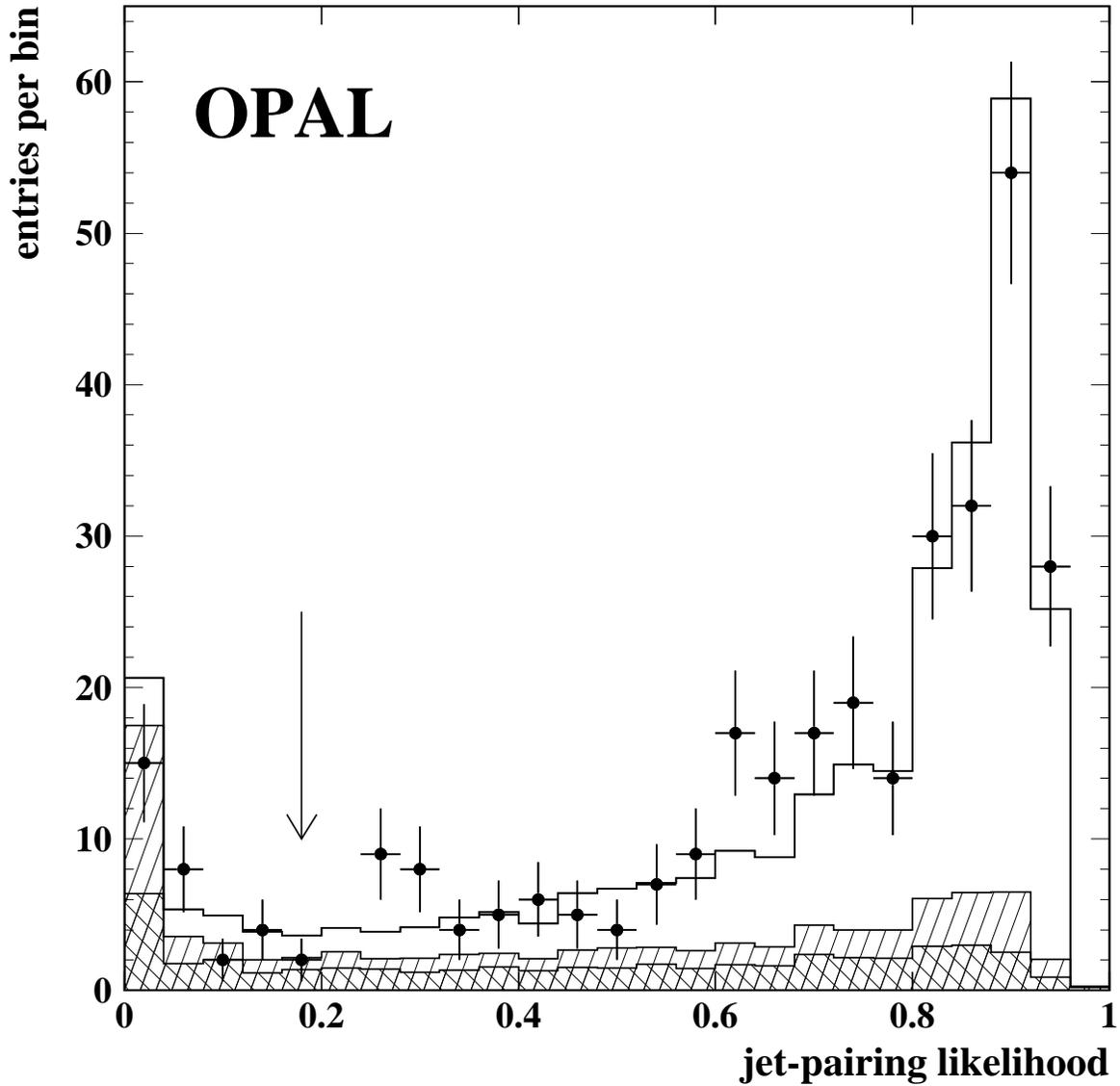}
  \caption{ The jet-pairing likelihood distribution for selected \WWqqqq\ 
    events.  For each event, only the likelihood output for the combination 
    yielding the maximum likelihood is plotted. Events to the right of 
    the arrow are retained for the mass analysis. The points correspond to the
    \Opal\ data and the open histogram to the Monte Carlo prediction. The 
    contribution from the non-WW background is shown as the cross-hatched 
    histogram and the addition of the combinatorial background is indicated by
    the singly-hatched histogram.}
\label{fig:jplh}
\end{figure}

For each \WWqqqq\ event three kinematic fits are performed, corresponding to 
the three possible jet pairings.  This ambiguity in the choice of the jet 
pairing leads to a combinatorial background.  To eliminate poorly reconstructed
events and reduce backgrounds, only combinations which yield a 5C fit with a 
$\chi^2$ fit-probability exceeding $0.01$ and $\mrec\: > \:65$~GeV are 
considered.  In addition, combinations with $\srec\:<\:0.5$~GeV are 
excluded\footnote{Fits which yield $\srec\:<\:0.5$~GeV are excluded because 
Monte Carlo studies reveal that the reconstructed mass resolution of these 
events is very poor; these events often have a reconstructed invariant mass 
close to the kinematic limit and are assigned an anomalously small fit error.}.
A relative likelihood discriminant is employed to choose a single combination 
for each event.  The likelihood is constructed for each surviving combination 
and takes as input the following three variables: the difference between the 
two fitted masses resulting from a 4C fit, the sum of the di-jet opening angles
and the 5C fit mass. The resulting jet-pairing likelihood distribution is shown
in Figure~\ref{fig:jplh}.  For each event, the combination corresponding to the
largest jet-pairing likelihood is retained provided it has a likelihood output 
exceeding $0.18$. If the combination fails to satisfy this criterion, the event
is not used. The cut of $0.18$ on the jet-pairing likelihood is chosen to 
optimize the product of efficiency and purity. Approximately $69\%$ of 
selected \WWqqqq\ events survive, while $60\%$ of the background is removed.  
Most of the rejected events fail the cut on the $\chi^2$ fit-probability.  
Monte Carlo studies estimate that in $87\%$ of the surviving signal events, the
selected combination corresponds to the correct jet pairing and that this 
fraction is independent of \Mw\ to within $0.1\%$ over a $\pm 1$~GeV range.  
The number of surviving events in the \WWqqqq\ channel is given in 
Table~\ref{tab:evno}.

In the selected \WWqqen\ and \WWqqmn\ events the non-leptonic part of the 
event is reconstructed as two jets using the Durham algorithm.   A kinematic 
fit is then performed incorporating the same five constraints as employed for 
the \WWqqqq\ events. This results in a 2C fit since the three-momentum of the 
neutrino is not known.  For the leptons, the inputs to the kinematic fit are 
the lepton energy and direction and their associated errors.  The direction is 
estimated using the track associated with the electron or muon candidate.  The
energy is estimated from the associated electromagnetic calorimeter cluster for
electrons and from the momentum of the associated track for muons. Jets are 
treated in the manner described above.  Only events with a $\chi^2$ 
fit-probability exceeding $0.001$, $\mrec\: > \:65$~GeV and 
$\srec\: >\: 0.5$~GeV are retained.  These cuts reduce backgrounds by roughly 
$55\%$ and remove poorly reconstructed events.  Since the \WWqqln\ event 
selections already yield low backgrounds, the cut on the $\chi^2$ 
fit-probability is looser than for the \WWqqqq\ channel.  Approximately $90\%$
and $92\%$ of selected \WWqqen\ and \WWqqmn\ events, respectively, satisfy 
these additional criteria.  Most of the rejected events fail the cut on the 
$\chi^2$ fit-probability.  The numbers of surviving events are listed in 
Table~\ref{tab:evno}.

The Selected \WWqqtn\ events are also reconstructed as two jets using the 
Durham algorithm after excluding the tracks and clusters associated with the 
tau.  The invariant mass of the jet-jet system, scaled by the ratio of the beam
energy to the sum of the jet energies, and its associated error are used in 
determining \Mw\ from this channel.  Only events with a reconstructed invariant
mass greater than $65$~GeV and an error on the reconstructed invariant mass 
greater than $0.5$~GeV are retained.  In addition, to further reduce 
background, a 1C fit is performed and the resulting $\chi^2$ fit-probability is
required to exceed $0.001$.  The fit incorporates the same five constraints as
employed for the \WWqqqq\ channel, assumes that the $\tau$ lepton direction is
given by the direction of the visible decay products associated with the tau 
and estimates the total energy of the tau using energy and momentum 
constraints.  Approximately $78\%$ of selected \WWqqtn\ events satisfy these 
additional criteria while $65\%$ of the background is removed.  Most of the 
rejected events fail the cut on the $\chi^2$ fit-probability.  The number of 
surviving events is listed in Table~\ref{tab:evno}.

The full width at half maximum of the residual of the reconstructed invariant 
mass per event is used as an estimate of the average \mrec\ resolution and 
is calculated using Monte Carlo events in which less than $100$~MeV of energy 
is radiated into initial state photons.  For \WWqqqq\ events this resolution 
is $1.7$~GeV for fits corresponding to the correct jet pairing.  For \WWqqln\ 
events the average \mrec\ resolution per event is $2.4$, $2.8$ and $3.4$~GeV 
in the \WWqqen, \WWqqmn\ and \WWqqtn\ channels, respectively. 

\subsection{Extraction of the W Mass and Width}
\label{ssec:RWfit}

The W boson mass and width are extracted by directly comparing the 
reconstructed mass distribution of the data to mass spectra obtained from fully
simulated Monte Carlo events corresponding to various values of \Mw\ and \Gw. A
likelihood fit is used to extract \Mw\ and \Gw\ by determining which Monte 
Carlo spectrum best describes the data.  The Monte Carlo spectra for arbitrary
values of \Mw\ and \Gw\ are obtained using the Monte Carlo reweighting 
technique described in~\cite{bib:O-mw172}.

\begin{figure}
  \epsfxsize=\textwidth
  \epsffile{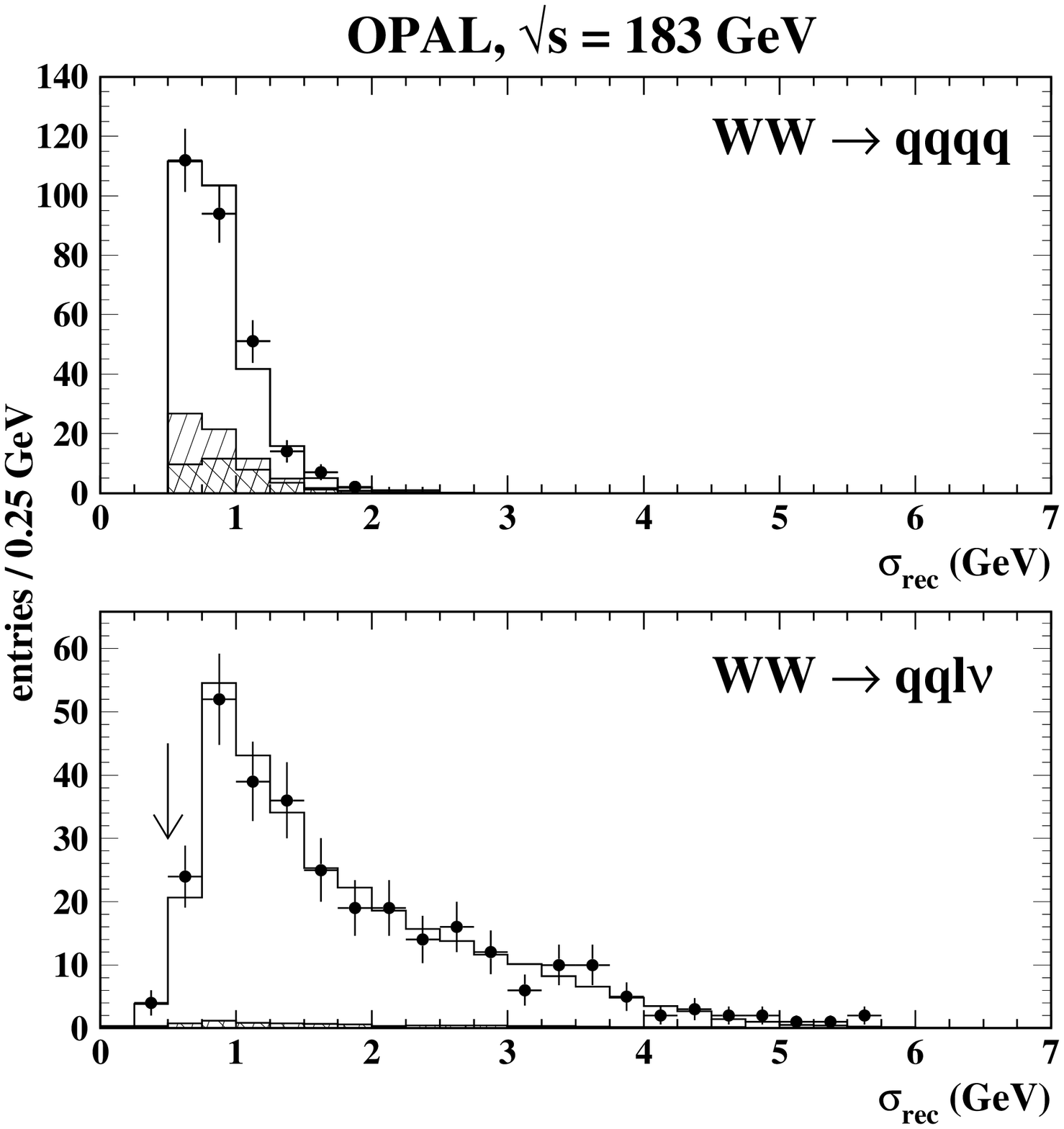}
  \caption{ The distribution of the error on the reconstructed invariant mass,
    \srec, for \WWqqqq\ events surviving the jet-pairing likelihood cut and 
    used in the mass analysis (top), and for all selected \WWqqln\ events 
    (bottom).  The \WWqqln\ events to the right of the arrow are retained for
    the mass analysis.  This error is used to divide the \WWqqln\ events into 
    four subsamples: $0.5 < \srec < 1.5$~GeV, $1.5 < \srec < 2.5$~GeV, 
    $2.5<\srec < 3.5$~GeV, $\srec > 3.5$~GeV, which are chosen to minimize the
    expected statistical error on \Mw.  The points correspond to the \Opal\ 
    data and the open histogram to the Monte Carlo prediction. The background 
    contribution is indicated by the cross-hatched histogram. In the \WWqqqq\ 
    channel, the addition of the combinatorial background is indicated by the 
    singly-hatched histogram.}
\label{fig:emf}
\end{figure}

The mass spectra for background events are taken from Monte Carlo and are 
assumed to be independent of \Mw\ and \Gw.  The background reconstructed mass 
distributions are normalised to the expected number of background events.  The
reweighted signal spectra are then normalised such that the total number of 
signal plus background events corresponds to the observed number of events. 
This is done separately for the \WWqqqq , \WWqqen , \WWqqmn\ and \WWqqtn\ 
channels.  In addition, the \WWqqln\ channels are divided into four subsamples 
according to the error on the reconstructed invariant mass. These subsamples
are treated independently within each channel.   This division 
gives a larger weight to events with reconstructed masses which are known with 
better precision (\ie\ small \srec ) relative to events with poorly determined
\mrec\ and reduces the expected statistical uncertainty on the fitted W mass
by approximately $7\%$ in the \WWqqln\ channels.   In the \WWqqqq\ channel, the
width of the reconstructed mass distribution is dominated by the intrinsic 
width of the W so that a similar subdivision does not improve the \Mw\ 
sensitivity in this channel and therefore is not implemented.  The \srec\ 
distribution is shown in Figure~\ref{fig:emf} for the \WWqqqq\ and \WWqqln\ 
channels separately.

A binned log-likelihood fit to the \mrec\ distributions of the data is 
performed in the range $\mrec > 65~\GeV$.  The log-likelihood function is 
defined as
\begin{displaymath}
  \ln(\mathcal{L}) =
  \sum_{i=1}^{N_{\mrm{bins}}}
  n_i \ln{( f_{b} \, \mathcal{P}_{b}^{i} 
       + (1-f_{b}) \, \mathcal{P}_{s}^{i}(\Mw,\Gw) )},  
\end{displaymath}
where $n_i$ is the number of observed events in the $i$th bin, $f_{b}$ is the
expected background fraction of the sample using the normalization procedure
described above, $\mathcal{P}_{s}^{i}(\Mw,\Gw)$ is the probability of 
observing a signal event in the $i$th bin assuming a W boson mass and width of
\Mw\ and \Gw\ and $\mathcal{P}_{b}^{i}$ is the analogous probability for the
background, which is assumed to be independent of the W mass and width.  Both 
$\mathcal{P}_{s}^{i}(\Mw,\Gw)$ and $\mathcal{P}_{b}^{i}$ are estimated using 
the relevant Monte Carlo spectrum. The log-likelihood curve is determined 
separately for each channel.  For the \WWqqln\ channels, the results are 
obtained by adding the log-likelihood curves separately determined from each 
subsample.

\begin{figure}
  \epsfxsize=\textwidth
  \epsffile{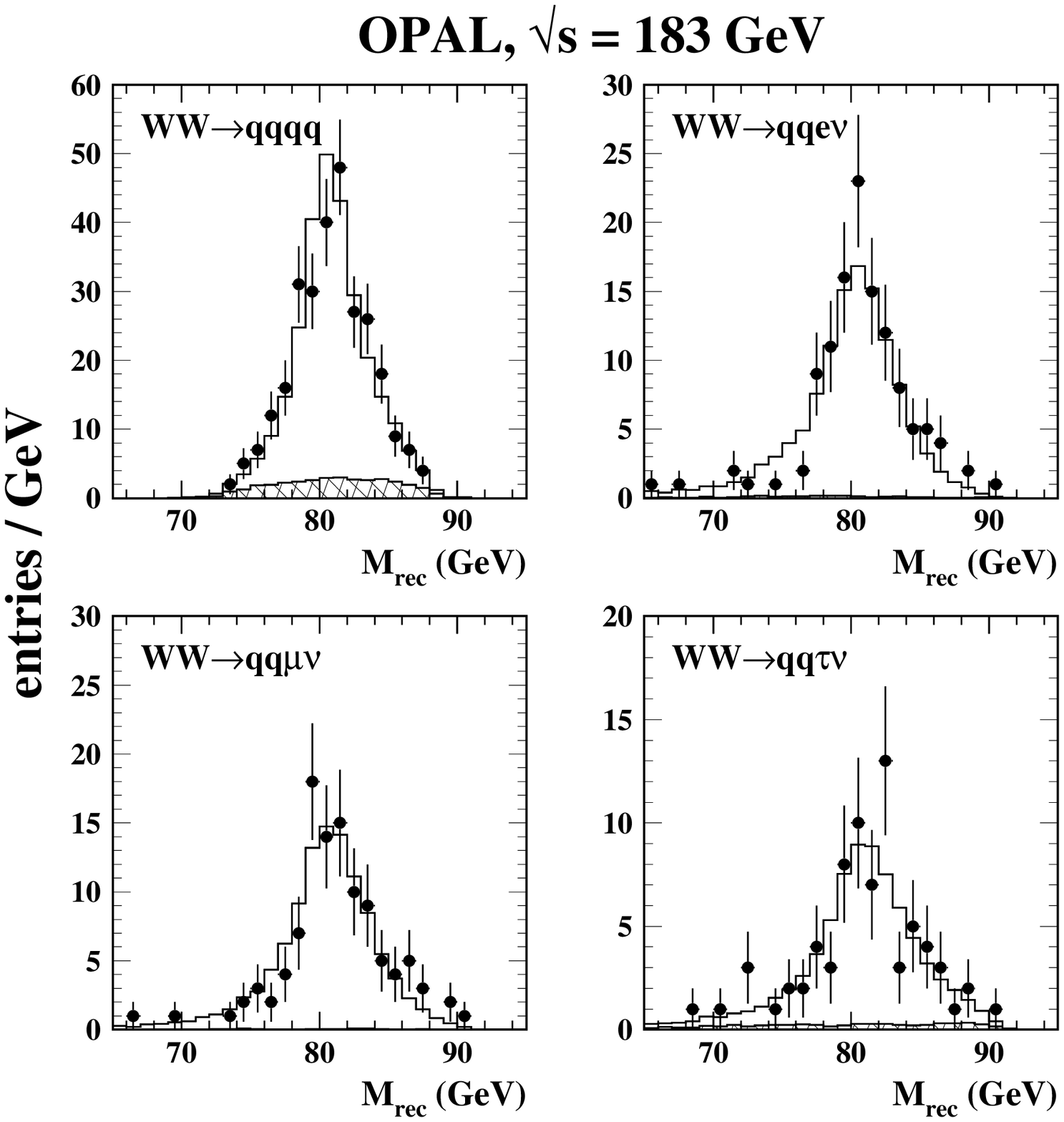}
  \caption{ The reconstructed invariant mass distribution for the \WWqqqq, 
    \WWqqen, \WWqqmn\ and \WWqqtn\ samples.  The points correspond to the
    \Opal\ data and the open histogram to the reweighted Monte Carlo spectrum
    corresponding to the fitted mass.  The background contribution is indicated
    by the cross-hatched histogram. }
\label{fig:rwdata}
\end{figure}

\begin{figure}[htb]
  \epsfxsize=\textwidth
  \epsfbox[0 0 567 680]{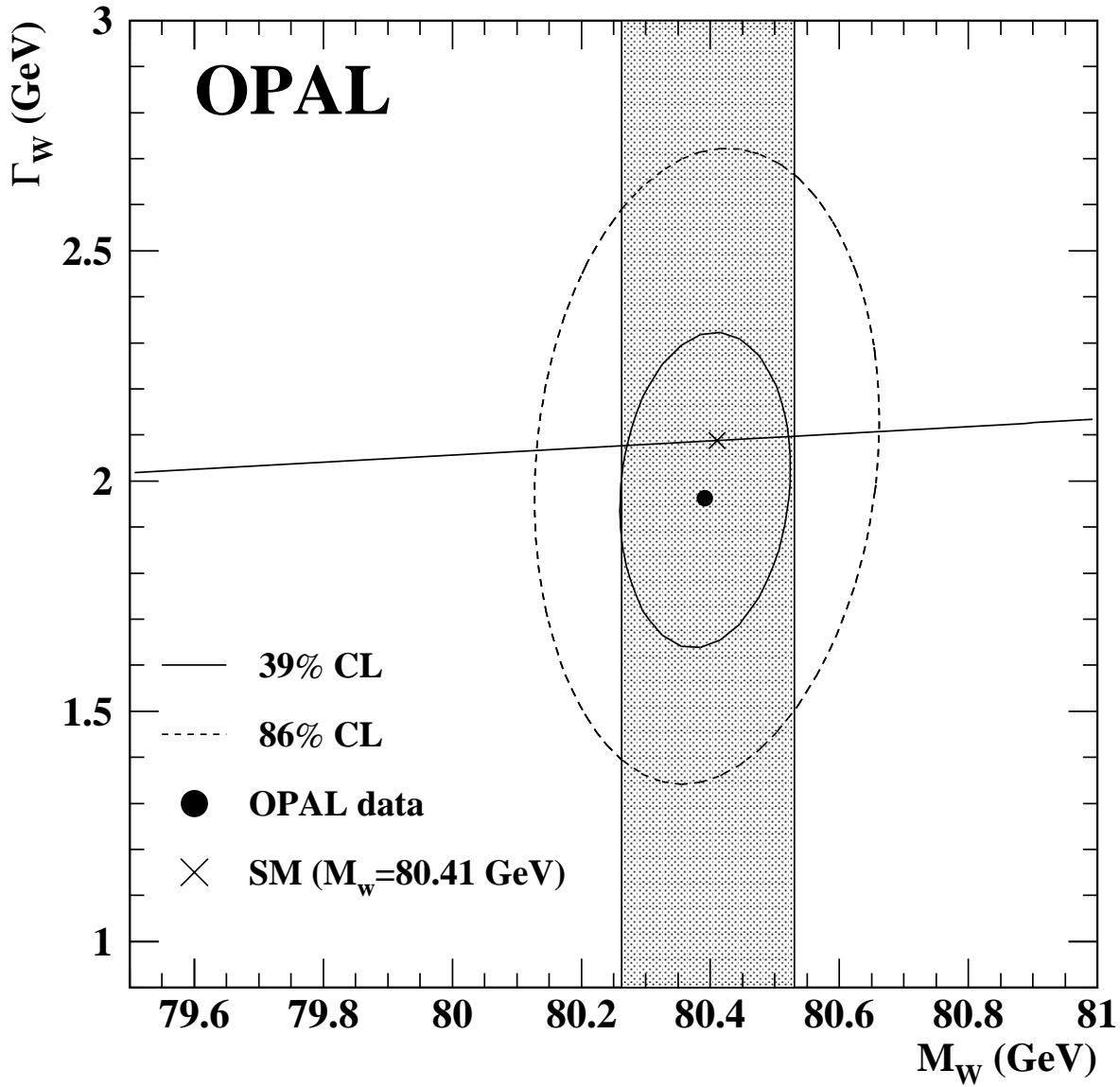}
  \caption
  {The 39\% and 86\% contour levels of the two-parameter fit using the
    reweighting method. The projections of these contours onto the axes give
    the one and two standard deviation statistical uncertainties.  The one 
    standard deviation region, including only the statistical error, of the fit
    to \Mw\ only is given by the shaded band. This fit is constrained to the 
    solid line, which gives the dependence of the width on the mass according 
    to the Standard Model.  The Standard Model prediction for \Gw, assuming the
    world average \Mw, is shown as an ``X''. }
 \label{fig:errcontour}
\end{figure}

Two types of fit are performed.  In the one-parameter fit, \Gw\ is constrained
to its Standard Model relation to the W mass \cite{bib:O-mw172} and only \Mw\
is determined.  The results of this fit for each channel are given in 
Table~\ref{tab:mw-rew} and displayed in Figure~\ref{fig:rwdata}.  The combined
result is discussed in Section~\ref{sec:results}.  In the two-parameter fit, 
both \Mw\ and \Gw\ are determined simultaneously.  The likelihood contours for
this fit are displayed in Figure~\ref{fig:errcontour} including statistical 
errors only.  The systematic uncertainties are discussed in 
Section~\ref{sec:syst}.

One advantage of the reweighting method is that the fitted parameters should
be unbiased since any offsets introduced in the analysis are implicitly 
accounted for in the Monte Carlo spectra used in the reweighting procedure.
This is verified using several Monte Carlo samples generated at various
\Mw\ and \Gw.  In addition, tests using a large ensemble of Monte Carlo 
subsamples, each corresponding to 57~pb$^{-1}$ and including background 
contributions, are used to verify for each channel separately and for all 
channels combined, that the measured fit errors accurately reflect the RMS 
spread of the residual distribution for both the \Mw\ and \Gw\ fits.  

\subsection{Alternative Fit Methods}
\label{sec:alt-fits}

\subsubsection{Breit-Wigner Fit}

The Breit-Wigner method is analogous to that described  in~\cite{bib:O-mw172}.
It employs an unbinned maximum-likelihood fit to the reconstructed mass 
spectrum using an analytic Breit-Wigner function to describe the signal. Due to
initial-state radiation,  the reconstructed mass spectrum is significantly 
asymmetric for data taken at $\roots\approx183$~GeV.  It is found that a 
relativistic Breit-Wigner function, with different widths above and below the 
peak, gives a satisfactory description of the \mrec\ lineshape. The function is
given by
\begin{displaymath}
  S( m_{\mrm{rec}} ) = A\frac{m^{2}_{\mrm{rec}} \, \Gamma_{-/+}^{2}}
   {(m^{2}_{\mrm{rec}}-m_{0}^{2})^{2} + m^{2}_{\mrm{rec}} \, \Gamma_{-/+}^{2}},
\end{displaymath}
where $\Gamma_{+(-)}$ is the width assumed for all \mrec\ above (below) the
peak, $m_{0}$.  The widths are fixed to values determined from fits to 
\WW\ signal Monte Carlo samples and are found to be independent of \Mw\ 
over the range relevant for this analysis.  The shapes of the background 
distributions and the background fraction are also determined from Monte Carlo.
The background fraction is held constant in the fit. The fit is performed over 
the range $70 < \mrec < 88$~GeV.

For the \WWqqqq\ events, the likelihood method for choosing which jet pairing
to use is found to distort the \mrec\ distribution so that it is inadequately
described by a Breit-Wigner function.  Therefore, the following procedure is 
employed \cite{bib:O-mw172}. The reconstructed invariant mass of the 
combination with the largest fit probability, $P_1$, is used if $P_1 > 0.01$.
The reconstructed invariant mass of the combination with the second largest fit
probability, $P_2$, is also used if it satisfies $P_2 > 0.01$ and 
$P_2 > P_1 / 3$.  The selected reconstructed masses enter the same distribution
with unit weight.  Monte Carlo studies estimate that the correct combination is
among those chosen in $90\%$ of the surviving \WWqqqq\ events.  

In contrast to the procedure employed for the reweighting method, the \WWqqln\
events are not divided into subsamples according to \srec\ because the 
subsamples exhibit a distorted reconstructed mass distribution which is poorly
described by a Breit-Wigner function. 

The fitted mass, $m_{0}$, must be corrected for offsets not accounted for in
the fit, \eg\ from initial-state radiation and event selection. A correction 
is determined using fully simulated Monte Carlo samples generated at 
various known \Mw\ and \roots\ with the expected background contributions 
included and is found to depend linearly on both \Mw\ and \roots.
The results from the \WWqqqq\ and \WWqqln\ channels, after correction, are
given in Table~\ref{tab:alt-fits}.

\subsubsection{Convolution Fit}

The convolution method is similar to that employed by the \Delphi\
Collaboration~\cite{bib:D-mw172}.  The method attempts to exploit all available
information by constructing a likelihood curve for each selected event.  The 
likelihood is calculated using the functional expression
\begin{displaymath}
  \mathcal{L}(\Mw , \mrec ) = p_s \mathcal{P}_{s}( \Mw , \mrec ), 
\end{displaymath}
where $p_s$ is the probability of the candidate event being a real signal
event and $\mathcal{P}_s$ is the probability density function for the signal,
\begin{displaymath}
  \mathcal{P}_s( \Mw , \mrec )=\mrm{BW}(\Mw , m ) \otimes \mrm{R}( m , \mrec ),
\end{displaymath}
where $\mrm{R}( m , \mrec )$ is the resolution function estimated from the 
Monte Carlo and $\mrm{BW}(\Mw , m )$ is a relativistic Breit-Wigner function
with $\Gw$ fixed to its Standard Model relation to $\Mw$.
The exact expression for the likelihood is channel dependent.  For example, in
the \WWqqqq\ channel the added complication of combinatorial background 
requires a sum over the three jet pairings.  The log-likelihood curves from
each selected event are summed to yield a single curve from which a fitted mass
is determined. This fitted mass is corrected for background effects and for 
offsets not accounted for in the fit in the manner described for the 
Breit-Wigner fit.  The results from the \WWqqqq\ and \WWqqln\ channels, after 
all corrections, are given in Table~\ref{tab:alt-fits}.

\section{Systematic Uncertainties}
\label{sec:syst}
The systematic uncertainties are estimated as described below and summarised in
Table~\ref{tab:sys}. 
\begin{description}
\item[Beam Energy:]\mbox{}\newline
  The average \Lep\ beam energy is known to a precision of $\pm 25$~MeV 
  \cite{bib:energy183}. The effect of this uncertainty on the measured \Mw\ is
  determined from fits to large Monte Carlo samples for which the analysis is
  repeated assuming $\roots = E^{\mrm{MC}}_{\mrm{cm}} \pm 50$~MeV.  The 
  observed shifts in the fitted \Mw\ are used to estimate the associated 
  systematic uncertainty.  The spread in \Lep\ beam energy of $152 \pm 8$~MeV
  \cite{bib:energy183} has a negligible effect on both the mass and width 
  determinations.
\item[Initial State Radiation:]\mbox{}\newline
  The systematic error associated with uncertainties in the modelling of 
  initial state radiation is estimated by comparing fully simulated \Koralw\ 
  Monte Carlo \WW\ events generated using a leading logarithm $O(\alpha)$ 
  treatment of initial state radiation to the standard \WW\ sample which 
  includes a next-to-leading-log $O(\alpha^2)$ treatment.  No significant 
  difference is observed and the statistical uncertainty of the comparison
  is taken as the associated systematic error.   
\item[Hadronization:]\mbox{}\newline 
  The scale of hadronization effects is studied by comparing the fit results
  of a single \WW\ sample generated once using \Pythia\ and again using
  \Herwig\ as the hadronization model. Both samples contain the same \WW\ final
  states and differ only in their hadronization modelling.  No significant 
  differences are observed and the statistical uncertainty of the comparison is
  taken as the associated systematic error.  As a cross-check, \WW\ 
  samples are generated with variations of the \Jetset\ fragmentation 
  parameters $\sigma_{q}$, $b$, $\Lambda_{\mrm{QCD}}$ and $Q_{0}$, of one 
  standard deviation about their tuned values \cite{jsettune}.  The fit results
  are compared and yield no statistically significant effects.
\item[Four-fermion Effects:]\mbox{}\newline
  The Monte Carlo samples used to estimate background contributions in the 
  reweighting procedure do not include a complete set of four-fermion diagrams
  and neglect interference effects between \WW\ diagrams and other 
  four-fermion processes. In order to test the sensitivity of the results to 
  these effects, the fit results of a sample generated including the full set 
  of interfering four-fermion diagrams are compared to those of a sample 
  restricted to the \CC\ \WW\ diagrams alone.  The comparison is 
  performed using both the \grcff\ and the \Excalibur\ generators.  In
  neither case is a significant difference observed.  The larger of the two 
  statistical uncertainties is assigned as the associated systematic error.
\item[Detector Effects:]\mbox{}\newline
   The effects of detector mis-calibrations and deficiencies in the Monte Carlo
   simulation of the data are investigated by varying the jet and lepton
   energy scales and the errors input to the kinematic fit over reasonable 
   ranges.  The ranges are dependent on polar angle and are determined from 
   detailed comparisons between 1997 data and Monte Carlo using approximately 
   $2.1\:\mathrm{pb}^{-1}$ of data collected at $\roots\approx\Mz$ and 
   $\epem\ra\epem$ events recorded at $\roots\approx 183$~GeV.
   Of particular importance for the \WWqqln\ channel are the lepton energy 
   scales, which are varied by $\pm 0.5\%$.  For the \WWqqqq\ channels the 
   most important variations are for the errors associated with the jet 
   angles, which are varied by $\pm 7\%$ for errors in $\cos\theta$ and 
   $\pm 3\%$ for errors in $\phi$.
   The jet energy scale is varied by $\pm 1$--$2\%$ over most of 
   $\left|\cos\theta\right|$.  For each variation a large Monte Carlo sample is
   refitted and the resulting shifts in the fitted \Mw\ are added in quadrature
   to yield an estimate of the associated systematic error.
\item[Fit Procedure:]\mbox{}\newline
   The reweighting procedure accounts for the fact that the Monte Carlo samples
   are generated at a centre-of-mass energy different from that of the data
   \cite{bib:O-mw172}. Monte Carlo samples generated at $\roots = 182$ and 
   $184$~GeV are used to test this procedure and assign a systematic error.

   The three methods used to measure \Mw\ each utilise the data and Monte Carlo
   in different ways.  A comparison of the fitted results is used to test for 
   residual biases in the reweighting fit.  The three methods are compared 
   using an ensemble of Monte Carlo subsamples, each corresponding to 
   $57$~pb$^{-1}$ and including background.  For each subsample the difference
   in the fitted \Mw\ determined using the reweighting method and that 
   determined using each of the other methods is calculated for the \WWqqqq\ 
   and \WWqqln\ channels separately and for the combined sample.  The 
   distributions of these differences are approximately Gaussian and have means
   consistent with zero and RMS values of approximately $90$~MeV, $95$~MeV, and
   $75$~MeV ($100$~MeV, $75$~MeV, and $60$~MeV) when comparing the reweighting
   fits to the Breit-Wigner (convolution) fits in the \WWqqqq, \WWqqln\ and the
   combined samples, respectively.  Since these alternative analyses yield 
   results consistent with those obtained using the default reweighting 
   analysis, no additional systematic is assigned based on these comparisons.
\item[Background Treatment:]\mbox{}\newline
   Uncertainties associated with both the normalization and shape of the 
   background distributions are investigated.  The background normalization is
   varied by one standard deviation of its associated uncertainties as 
   evaluated in~\cite{bib:O-tgc183} and the data are refitted.  As an estimate
   of the errors associated with the uncertainties in the shape of the 
   background distributions, a variety of substitutions are made.  For all 
   channels, \Herwig\ replaced \Pythia\ for the hadronization model for the 
   \Zqq\ background.  In addition, for the \WWqqqq\ channel, data taken at 
   $\roots\approx\Mz$, scaled by $( 183\:\GeV / \Mz )$, are also substituted 
   for the \Zqq\ background.  For each substitution the data are refitted. The 
   quadrature sum of the shift in fitted mass observed when changing the 
   normalization and the largest of the observed shifts in the fitted mass from
   the various substitutions in each channel is assigned as a systematic error.
   Using \Pythia\ samples of \Wenu\ events generated using various \Mw, it is 
   verified that the uncertainty on \Mw\ has a negligible effect on the 
   background distributions.
\item[Monte Carlo Statistics:]\mbox{}\newline
   The finite statistics of the Monte Carlo samples used in the 
   reweighting procedure contribute a systematic uncertainty of $\pm 15$~MeV 
   to the W mass determined separately in the \WWqqqq\ and \WWqqln\ channels, 
   and $\pm40$~MeV to the W width determined from the combined sample.
\item[Colour-Reconnection Effects and Bose-Einstein Correlations:]\mbox{}\newline
  As discussed in \cite{bib:LEP2YR} and \cite{bib:wwqcd} and references 
  therein, a significant bias to the apparent W mass measured in the \WWqqqq\ 
  channel could arise from the effects of colour-reconnection and/or 
  Bose-Einstein correlations between the decay products of the \PWp\ and \PWm.
  These effects are investigated separately.

  Using currently available \WW\ data, it is not possible to discern whether
  or not Bose-Einstein correlations are present between hadrons originating 
  from different W decays [25-26].
%  from different W decays \cite{bib:O-bec183}--$\!\!$\cite{bib:D-bec183}.
  To investigate possible systematic biases a Monte Carlo sample is generated 
  including Bose-Einstein correlations using the \Pythia\ Monte Carlo 
  generator implemented as described in \cite{bib:pyt-be}.  The fit result in
  the \WWqqqq\ channel from this sample is compared to a fit from a \Pythia\ 
  sample which excludes Bose-Einstein correlations.  No significant bias is 
  observed for the fitted parameters and the statistical precision of the 
  comparison is taken as the associated systematic error of $\pm 32$~MeV in the
  fitted mass determined from the \WWqqqq\ channel.  The uncertainty on the
  fitted width in \WWqqqq\ channel propagates to an uncertainty of $\pm 55$~MeV
  on the fitted width determined from the combined sample.  

  To investigate the systematic biases originating from colour-reconnection
  effects, several models are studied using the \Pythia\ and \Ariadne\ 
  \cite{bib:ard} Monte Carlo generators.  As discussed in the accompanying 
  paper \cite{bib:O-cr183}, 
  comparing various event shapes offers a means of discriminating between the 
  models and, when comparing with data, a means of testing each model 
  independently of the \Mw\ measurement.  Based on these studies, and
  comparing fitted masses between Monte Carlo samples including and excluding
  colour-reconnection effects, a systematic uncertainty of $\pm 49$~MeV is 
  assigned to the mass determined from the \WWqqqq\ channel \cite{bib:O-cr183}.
  The uncertainty on the fitted width in \WWqqqq\ channel propagates to an 
  uncertainty of $\pm 109$~MeV on the fitted width determined from the 
  combined sample.  
\end{description}

\noindent
The contributions from each of the above sources are added in quadrature to
yield the total systematic uncertainty.  For the alternative analyses, the 
systematics are estimated similarly and yield comparable results.

\section{Results}
\label{sec:results}

For the reweighting method described in Section~\ref{ssec:RWfit}, the results 
of a simultaneous fit to \Mw\ and \Gw\ from the combined \WWqqqq\ and \WWqqln\ 
event samples are
\begin{eqnarray*}
 \Mw & = & 80.39 \pm 0.13 \pm 0.06 ~\GeV, \\
 \Gw & = &  1.96 \pm 0.34 \pm 0.20 ~\GeV,
\end{eqnarray*}
where the uncertainties are statistical and systematic, respectively.  The 
correlation coefficient between \Mw\ and \Gw\ is 0.13.  For this fit, the
central values are determined by adding the log-likelihood curves from the
\WWqqqq\ and \WWqqln\ channels.  The systematic uncertainties are also 
estimated by summing the log-likelihood curves from each channel.

A one-parameter fit for the mass is performed by constraining the width using 
the Standard Model relation to give 
$\Mw =  80.53 \pm 0.23 (\mrm{stat}) \pm 0.09 (\mrm{syst})$~GeV in the \WWqqqq\
channel, and $\Mw = 80.33 \pm 0.17 (\mrm{stat}) \pm 0.06 (\mrm{syst})$~GeV in
the \WWqqln\ channel.  The combined result is determined taking into account 
the correlated systematics between the \WWqqqq\ and \WWqqln\ channels and gives
\begin{eqnarray*}
   \Mw & = & 80.39 \pm 0.13 \pm 0.05 ~\GeV,
\end{eqnarray*}
where the uncertainties are statistical and systematic, respectively.  For the
combination, the \WWqqqq\ channel carries a weight of $0.34$.
The combined \WWqqqq\ and \WWqqln\ results from the alternative analyses, after
all corrections, are for the Breit-Wigner fit, 
$\Mw=80.37\pm 0.15 (\mrm{stat}) \pm 0.05 (\mrm{syst})$~GeV, and for the 
convolution fit, $\Mw=80.30\pm 0.14 (\mrm{stat}) \pm 0.06 (\mrm{syst})$~GeV.  
As discussed in Section~\ref{sec:syst}, these results are statistically 
consistent with those obtained using the reweighting fit.

The difference between the fitted \Mw\ in the \WWqqqq\ and \WWqqln\ channels is
$\DMw \equiv ( \Mwqqqq - \Mwqqln ) = 0.20 \pm 0.28 \pm 0.07 ~\GeV $, where
the uncertainties are statistical and systematic (excluding contributions from
colour-reconnection/Bose-Einstein effects), respectively.  A significant 
non-zero value for \DMw\ could indicate that colour-reconnection/Bose-Einstein
effects are biasing the \Mw\ determined from \WWqqqq\ events.

\subsection{Combination with Previous Data}

The measurements of \Mw\ from direct reconstruction at $\roots\approx172$~GeV 
\cite{bib:O-mw172} and $\roots\approx183$~GeV are combined with the \Mw\ 
measurement from the \WW\ production cross-section at threshold, 
$\roots\approx161$~GeV \cite{bib:O-mw161}. The combination is made assuming 
that the mass measurements from direct reconstruction and from the threshold 
cross-section are uncorrelated, apart from the uncertainty associated with the
\Lep\ beam energy, which is taken to be fully correlated.  The direct 
reconstruction measurements are combined accounting for correlated systematics.
The combined result is
\begin{displaymath}
        \Mw = 80.38 \pm 0.12 \pm 0.04 \pm 0.02 \pm 0.02 \, \GeV,
\end{displaymath} 
where the uncertainties are statistical, systematic, 
colour-reconnection/Bose-Einstein and beam energy, respectively.

The measurements of \Gw\ from direct reconstruction at $\roots\approx172$~GeV
\cite{bib:O-mw172} and $\roots\approx183$~GeV are combined taking into account
the correlated systematics to obtain
\begin{displaymath}
        \Gw = 1.84 \pm 0.32 \pm 0.15 \pm 0.12 \pm 0.01 \, \GeV,
\end{displaymath} 
where the uncertainties are statistical, systematic, 
colour-reconnection/Bose-Einstein and beam energy, respectively.  Monte Carlo 
studies reveal that the measured statistical error is correlated with the 
measured width.  To avoid biasing the combination, the separate width 
measurements are weighted using the expected statistical error, determined from
an ensemble of many Monte Carlo experiments assuming a W width of $2.093$~GeV.

The measurements of \DMw\ from direct reconstruction at $\roots\approx172$~GeV
and $\roots\approx183$~GeV are combined taking into account the correlated 
systematics to obtain
\begin{displaymath}
        \DMw = 0.08 \pm 0.26 \pm 0.07 \, \GeV,
\end{displaymath} 
where the uncertainties are statistical and systematic, respectively. Note 
that no systematic error is included for uncertainties in the modelling of 
colour-reconnection/Bose-Einstein effects and that the uncertainty in the 
\Lep\ beam energy does not contribute a systematic error to this quantity.

\section{Summary}

Using the $57$~pb$^{-1}$ of data recorded by the \Opal\ detector at a mean 
centre-of-mass energy of approximately 183~GeV, a total of 582 \WWqqqq\ and 
\WWqqln\ candidate events are used in a fit constraining \Gw\ to its Standard 
Model relation with \Mw\ to obtain a direct measurement of the W boson mass,
{\ensuremath{\Mw=80.39\pm0.13\mrm{(stat.)}\pm0.05\mrm{(syst.)}}}~GeV, while
a second fit is used to directly determine the width of the W boson,
{\ensuremath{\Gw=1.96\pm0.34 \mrm{(stat.)}\pm0.20\mrm{(syst.)}}}~GeV.

The results described in this paper are combined with the previous \Opal\ 
results from data recorded at $\roots \approx 172$~GeV and 
$\roots\approx161$~GeV.  From this combined data sample the W boson mass is
determined to be
\begin{displaymath}
        \Mw = 80.38 \pm 0.12 \pm 0.05\, \GeV.
\end{displaymath} 
The result for the W boson width is combined with the previous \Opal\ result 
from data recorded at $\roots \approx 172$~GeV to obtain
\begin{displaymath}
        \Gw = 1.84 \pm 0.32 \pm 0.20\, \GeV.
\end{displaymath} 
The uncertainties are statistical and systematic, respectively.  

% --- pigpigpig
%
% --- acknowledgements
%
\clearpage
\appendix
\noindent
Acknowledgements:
\par
\noindent
We particularly wish to thank the SL Division for the efficient operation
of the LEP accelerator at all energies
 and for their continuing close cooperation with
our experimental group.  We thank our colleagues from CEA, DAPNIA/SPP,
CE-Saclay for their efforts over the years on the time-of-flight and trigger
systems which we continue to use.  In addition to the support staff at our own
institutions we are pleased to acknowledge the  \\
Department of Energy, USA, \\
National Science Foundation, USA, \\
Particle Physics and Astronomy Research Council, UK, \\
Natural Sciences and Engineering Research Council, Canada, \\
Israel Science Foundation, administered by the Israel
Academy of Science and Humanities, \\
Minerva Gesellschaft, \\
Benoziyo Center for High Energy Physics,\\
Japanese Ministry of Education, Science and Culture (the
Monbusho) and a grant under the Monbusho International
Science Research Program,\\
Japanese Society for the Promotion of Science (JSPS),\\
German Israeli Bi-national Science Foundation (GIF), \\
Bundesministerium f\"ur Bildung, Wissenschaft,
Forschung und Technologie, Germany, \\
National Research Council of Canada, \\
Research Corporation, USA,\\
Hungarian Foundation for Scientific Research, OTKA T-016660, 
T023793 and OTKA F-023259.\\

%
% --- References
%

%
% --- Tables
%
\clearpage

\renewcommand{\arraystretch}{1.0}
\begin{table}[htbp]
  \begin{center}
    \begin{tabular}{|l|ccc|} \hline
      Channel & Observed & Expected & Purity \\ \hline
      \WWqqqq\       & 282 & 278.1 & $86\%$ \\
      \WWqqenu\      & 119 & 113.0 & $97\%$ \\
      \WWqqmnu\      & 107 & 114.4 & $99\%$ \\
      \WWqqtnu\      &  74 &  82.6 & $95\%$ \\ 
      Combined       & 582 & 588.0 & $92\%$ \\ \hline
    \end{tabular}
  \end{center}
  \caption[foo]{
    Numbers of events used in the W mass and width determination for each 
    channel and all channels combined.  Only events surviving the cuts 
    described in Section~\ref{mrec} are included.  The number of expected 
    events and corresponding purities are estimated assuming the world 
    average \Mw\ and have relative uncertainties of approximately $3\%$, 
    dominated by the uncertainty in the \CC\ production cross-section.}
  \label{tab:evno}
\end{table} 
\vspace*{1.0cm}

\begin{table}[htbp]
  \begin{center}
    \begin{tabular}{|l|cc|} \hline
      Channel & Measured \Mw  & Expected error \\ \hline
      \WWqqen & $80.23 \pm 0.24$ & $0.31$ \\
      \WWqqmn & $80.42 \pm 0.27$ & $0.30$ \\
      \WWqqtn & $80.40 \pm 0.40$ & $0.45$ \\ \hline
      \WWqqln & $80.33 \pm 0.17$ & $0.19$ \\
      \WWqqqq & $80.53 \pm 0.23$ & $0.20$ \\ \hline
    \end{tabular}
  \caption{ Results using the reweighting method for the fit constraining \Gw\
    to its Standard model relation with \Mw\ from 
    $57$~pb$^{-1}$ of data taken at $\roots\approx183$~GeV for each of the 
    channels separately and for the combined \WWqqln\ channel. The expected
    errors are estimated using an ensemble of Monte Carlo subsamples, each
    corresponding to $57$~pb$^{-1}$ of data and including background 
    contributions.  The errors are statistical only.  (All quantities are in 
    GeV.)}
  \label{tab:mw-rew}
  \end{center}
\end{table}
\vspace*{1.0cm}

\begin{table}[htbp]
  \begin{center}
    \begin{tabular}{|l|c|c|} \hline
        & Breit-Wigner fit & Convolution fit \\
      Channel & Measured \Mw     & Measured \Mw     \\ \hline
      \WWqqln & $80.27 \pm 0.19$ & $80.24 \pm 0.19$ \\
      \WWqqqq & $80.52 \pm 0.24$ & $80.38 \pm 0.21$ \\ \hline
    \end{tabular}
  \caption{ Fit results using the alternative analyses and $57$~pb$^{-1}$ 
    of data taken at $\roots\approx183$~GeV for the \WWqqln\ and \WWqqqq\ 
    channels separately.  The expected statistical errors are very similar to
    those of the reweighting method given in Table~\ref{tab:mw-rew}.  The
    errors are statistical only.  (All quantities are in GeV.)}
  \label{tab:alt-fits}
  \end{center}
\end{table}
\vspace*{1.0cm}

\begin{table}[htbp]
  \begin{center}
    \begin{tabular}{|l||c|c|c||c||c|} \hline
      Systematic errors &\multicolumn{3}{|c||}{ \Mw } & &     \\
        ~~~~~~~~(MeV) 
      & \qqqq & \qqln & comb. & $\Gw$ & $\DMw$                \\ \hline
      Beam Energy                & 22  & 22 & 22 & 5   &  0   \\
      Initial State Radiation    & 10  & 10 & 10 & 15  & 10   \\
      Hadronization              & 21  & 21 & 16 & 52  & 15   \\
      Four-fermion               & 30  & 28 & 21 & 52  & 40   \\
      Detector Effects           & 38  & 31 & 26 & 90  & 46   \\
      Fit Procedure              & 15  & 15 & 15 & 45  & 20   \\
      Background                 & 25  & 10 & 10 & 79  & 25   \\
      MC statistics              & 15  & 15 & 11 & 40  & 21   \\ \hline
      Sub-total                  & 67  & 58 & 49 & 154 & 74   \\
      Bose-Einstein Correlations & 32  &  0 & 11 & 55  & ---  \\ 
      Colour Reconnection        & 49  &  0 & 16 & 109 & ---  \\ \hline\hline
      Total systematic error     & 89  & 58 & 53 & 196 & 74   \\ \hline
    \end{tabular}
  \end{center}
  \caption[foo]{
    Summary of the systematic uncertainties for the fit results.  For the fits
    to determine \Mw, \Gw\ is constrained to its Standard Model relation.
    The uncertainties are given separately for fits to the \WWqqqq, \WWqqln\ 
    and the combined samples.  For \Gw\ the uncertainties are given only for 
    the fit to the combined sample.  The quantity 
    $\DMw\equiv\left(\Mwqqqq - \Mwqqln \right)$ uses the Standard Model 
    constrained fits to the \WWqqqq\ and \WWqqln\ channels separately.  Since
    the primary interest in the quantity \DMw\ is as a test of possible
    colour-reconnection/Bose-Einstein effects in the \WWqqqq\ channel, no 
    systematic error is assigned for uncertainties associated with the 
    modelling of these effects. The inter-channel correlations are taken into
    account for the combined \Mw\ and \Gw\ fits and for the \DMw\ 
    determination.}
  \label{tab:sys}
\end{table} 
\vspace*{1.0cm}

%%%\begin{table}[htbp]
%%%  \begin{center}
%%%    \begin{tabular}{|l|r|} \hline
%%%      ~~~~~~~Model & Mass shift (MeV) \\ \hline
%%%      ~~~~~~~~SK I            & $ 11\pm 25$~~~~~~ \\
%%%      ~~~~~~~~SK II           & $-23\pm 25$~~~~~~ \\
%%%      ~~~~~~~~SK II$^\prime$  & $-22\pm 25$~~~~~~ \\
%%%      ~~~~~~~~AR 2            & $ 49\pm 14$~~~~~~ \\
%%%      ~~~~~~~~AR 3            & $145\pm 21$~~~~~~ \\ \hline
%%%    \end{tabular}
%  \caption{ The predicted shifts in the fitted mass from the \WWqqqq\ channel
%  for various colour-reconnection (CR) models.  The shift is calculated as 
%  $M_{\mrm{fit}}(\mrm{CR}) - M_{\mrm{fit}}(\mrm{no\:CR})$ using two samples
%  generated with the same Monte Carlo program.  For the SK models 
%  $M_{\mrm{fit}}(\mrm{no\:CR})=80.360\pm0.018$~GeV and for the AR models
%  $M_{\mrm{fit}}(\mrm{no\:CR})=80.323\pm0.010$~GeV. All samples were generated
%  with $\Mw = 80.33$~GeV.  The \Ariadne\ 3 model, AR 3, is disfavoured in the 
%  data. The errors are statistical only.}
%%%  \label{tab:cr}
%%%  \end{center}
%%%\end{table}
%
% --- Figures
%
\newpage

\end{document}